\documentclass[10pt,a4paper]{scrartcl}
\usepackage[utf8]{inputenc}
\usepackage[english]{babel}
\usepackage{mathpazo}
\usepackage{a4wide}
\usepackage{amsmath,amssymb,bm}
\usepackage{tikz}
\usepackage{verbatim}
\definecolor{darkred}{rgb}{0.5,0,0}
\definecolor{darkblue}{rgb}{0,0,0.5}
\definecolor{firebrick}{rgb}{0.75,0.125,0.125}
\definecolor{darkgreen}{rgb}{0,0.5,0}
\usepackage[colorlinks=true,linkcolor=firebrick,citecolor=darkgreen,urlcolor=darkblue,pdfencoding=unicode]{hyperref}
\usepackage{xspace}
\usepackage{authblk}
\usepackage{cleveref}

\newcommand{\ngc}{\textsc{ngC}\xspace}
\def\Offline{\mbox{$\overline{\textrm{Off}}$\hspace{.05em}\protect\raisebox{.4ex}{$\protect\underline{\textrm{line}}$}}\xspace}

\title{Towards a Next Generation of CORSIKA:\\
A Framework for the Simulation of\\
Particle Cascades in Astroparticle Physics}

\author[1,2]{Ralph Engel}
\author[1]{Dieter Heck}
\author[1,3]{Tim Huege}
\author[1]{Tanguy Pierog}
\author[1,2]{Maximilian Reininghaus}
\author[4]{Felix Riehn}
\author[1]{Ralf Ulrich\thanks{e-mail: \href{mailto:ralf.ulrich@kit.edu}{ralf.ulrich@kit.edu}}}
\author[1]{Michael Unger}
\author[1]{Darko Veberič}
\affil[1]{Institut für Kernphysik, Karlsruher Institut für Technologie (KIT), Karlsruhe, Germany}
\affil[2]{Institut für Experimentelle Teilchenphysik, Karlsruher Institut für Technologie (KIT), Karlsruhe, Germany}
\affil[3]{Vrije Universiteit Brussel (VUB), Brussels, Belgium}
\affil[4]{Laboratório de Instrumentação e Física Experimental de Partículas (LIP), Lisboa, Portugal}

\date{August 2018}

\makeatletter
\hypersetup{
  pdftitle={Towards a Next Generation of CORSIKA},
  pdfauthor={Ralph Engel, Dieter Heck, Tim Huege, Tanguy Pierog, Maximilian Reininghaus, Felix Riehn, Ralf Ulrich, Michael Unger, Darko Veberič},
  pdflang=en
}
\makeatother

\usepackage[backend=biber,
  articletitle=true,
  style=phys,
  giveninits=true,
  biblabel=brackets,
  chaptertitle=false,
  pageranges=false,
  doi=true,
  eprint=true
]{biblatex}

\DeclareSourcemap{
        \maps[datatype=bibtex]{
                \map{
                        \step[fieldsource=Collaboration, match=\regexp{([^,]+)}, fieldset=usera, fieldvalue={$1}, final=true]
                }
        }
}
\renewbibmacro*{author}{%
        \iffieldundef{usera}{%
                \printnames{author}%
        }{%
        \printnames{author} (\printfield{usera})%
}%
}%
\addtokomafont{disposition}{\rmfamily}

\begin{document}

\maketitle

\begin{abstract}
A large scientific community depends on the precise modelling of
complex processes in particle cascades in various types of matter.  These models are used most
prevalently in cosmic-ray physics, astrophysical-neutrino
physics, and gamma-ray astronomy.  In this white paper, we summarize the necessary steps to ensure
the evolution and future availability of optimal simulation tools.  The
purpose of this document is not to act as a strict blueprint for next-generation software, but
to provide guidance for the vital aspects of its design.  The
topics considered here are driven by physics and scientific applications. Furthermore,
the main consequences of implementation decisions on performance are outlined.
We highlight the computational performance as an important aspect guiding the
design since future scientific applications will heavily depend on an efficient
use of computational resources.
\end{abstract}

\section{Introduction, History, and Context}

Simulations of air showers are an essential instrument for successful analysis
of cosmic-ray data.  The air-shower simulation program
CORSIKA~\cite{Heck:1998vt} is 
the leading tool for the research in this field. It has found use in many
applications, from calculating inclusive particle fluxes to simulating
ultra-high energy extensive air showers, and has been in the last decades
employed by most of the experiments (see e.g.~\cite{Horandel:2005at} and
references therein).
It has supported and helped shape the research during the last
25\,years with great success.  Originally designed as a FORTRAN\,77
program and as a part of the detector simulation for the KASCADE experiment
(the name itself comes from ``COsmic Ray SImulations for KAscade''), it was
soon adapted by other collaborations to their uses. The first were the
MACRO~\cite{Ambrosio:2002mb} and HEGRA~\cite{Daum:1997fp} experiments in 1993.
As a consequence, over the time it has evolved enormously and is nowadays used
by essentially all cosmic-ray, gamma-ray, and neutrino astronomy experiments.
Furthermore, it helped to create a universal common reference for the
worldwide interpretation and comparison of cosmic-ray air-shower
data. Before CORSIKA, it was very difficult for many types of
experiments to assess the physics content of their data, and almost
impossible to qualify the compatibility with different
measurements. 
In general, the simulation of extensive air showers was recognized as one of
the fundamental prerequisites for successful research in astroparticle
physics~\cite{Knapp:2002vs}.  In the past, some other tools have also been
developed for these purposes, of which the most well known are
MOCCA~\cite{Hillas:1997tf}, AIRES~\cite{Sciutto:1999jh} (with the extension
TIERRAS~\cite{Tueros:2010zz} for simulations of showers below ground), and
SENECA~\cite{Drescher:2002cr}.

Over all the years CORSIKA evolved into a large and hard to maintain example of
highly complex software, mostly due to the language features and restrictions
inherent to FORTRAN\,77.  While the performance is still excellent and the
mainstream use-cases are frequently tested as well as verified, it is
increasingly difficult to keep the development up-to-date with requests and
requirements.  It is becoming obvious that the limited features of the FORTRAN
language and the evident complexity of the new developments are getting into a
conflict. Furthermore, in the future, the expertise needed to maintain such a
large FORTRAN codebase will be more-and-more difficult to provide.
Therefore, it is important to make CORSIKA competitive for the challenges we
are facing in the future, requiring us to make a major step in terms of used
software technology. This will ensure that CORSIKA will evolve further and
become the most comprehensive and useful tool for simulating extensive particle
cascades in all required environments.

\section{Purpose and Aim}

The purpose of CORSIKA is to perform a ``particle transport with stochastic and
continuous processes''. A \emph{next-generation CORSIKA} (\ngc) will implement
this core task in the most direct, flexible, and efficient way. In this
document we will refer to this project as \ngc, but just as a simplification
and to clearly distinguish it from the existing CORSIKA program. The \ngc will
provide a framework where users can implement plugins and extensions for an
unspecified number of scientific problems to come. CORSIKA will take a step
from being an air-shower simulation program only, to becoming the most versatile
framework for particle-cascade simulations available.

The \ngc must support particle tracking, cascade equations (CE),
thinning, various particle interaction models, output options,
(massively) parallel computations including GPU support, various
possibilities for user routines to interact with the simulation
process, and full exposure of particles while they are
tracked/simulated. In particular the excellent performance of thinning
is critical for simulations at the highest energies~\cite{Risse:2002yd,Kobal:2001jx,Billoir:2008zz}. With \ngc it
will be possible to study thinning very precisely with techniques
known as \emph{multi-thinning}~\cite{Bruijn:2011}, in combination with a deep analysis of
the cascade history. It is important to improve the thinning
performance and technology with respect to the solutions available so
far.
Furthermore, production of Cherenkov photons, radio
signals, and similar non-cascade extensions should be fully supported.  As
usual, the cascades could be simulated in the atmosphere, but options for other
media or a combination of them will be added.

We expect that millions, if not billions, of CPU hours of high
performance computing will be spent in the future on air-shower simulations for experiments
like CTA~\cite{Consortium:2010bc}, H.E.S.S.~\cite{Hinton:2004eu}, IceCube~\cite{Achterberg:2006md},
LOFAR~\cite{vanHaarlem:2013dsa}, MAGIC~\cite{Aleksic:2014lkm}, the Pierre Auger
Observatory~\cite{Abraham:2004dt}, the Telescope Array~\cite{AbuZayyad:2012kk},
and other next-generation experiments. It is
up to \ngc to make sure this is done as efficiently and accurately as possible,
while maximising the resulting physics output. In this respect, \ngc plays an
important role in spending valuable and sparse resources while it is at the
same time a fundamental cornerstone supporting the physics output of many large
experiments.

\section{Main design considerations}

\begin{figure}[t!]
  \centering
  \includegraphics[width=\textwidth]{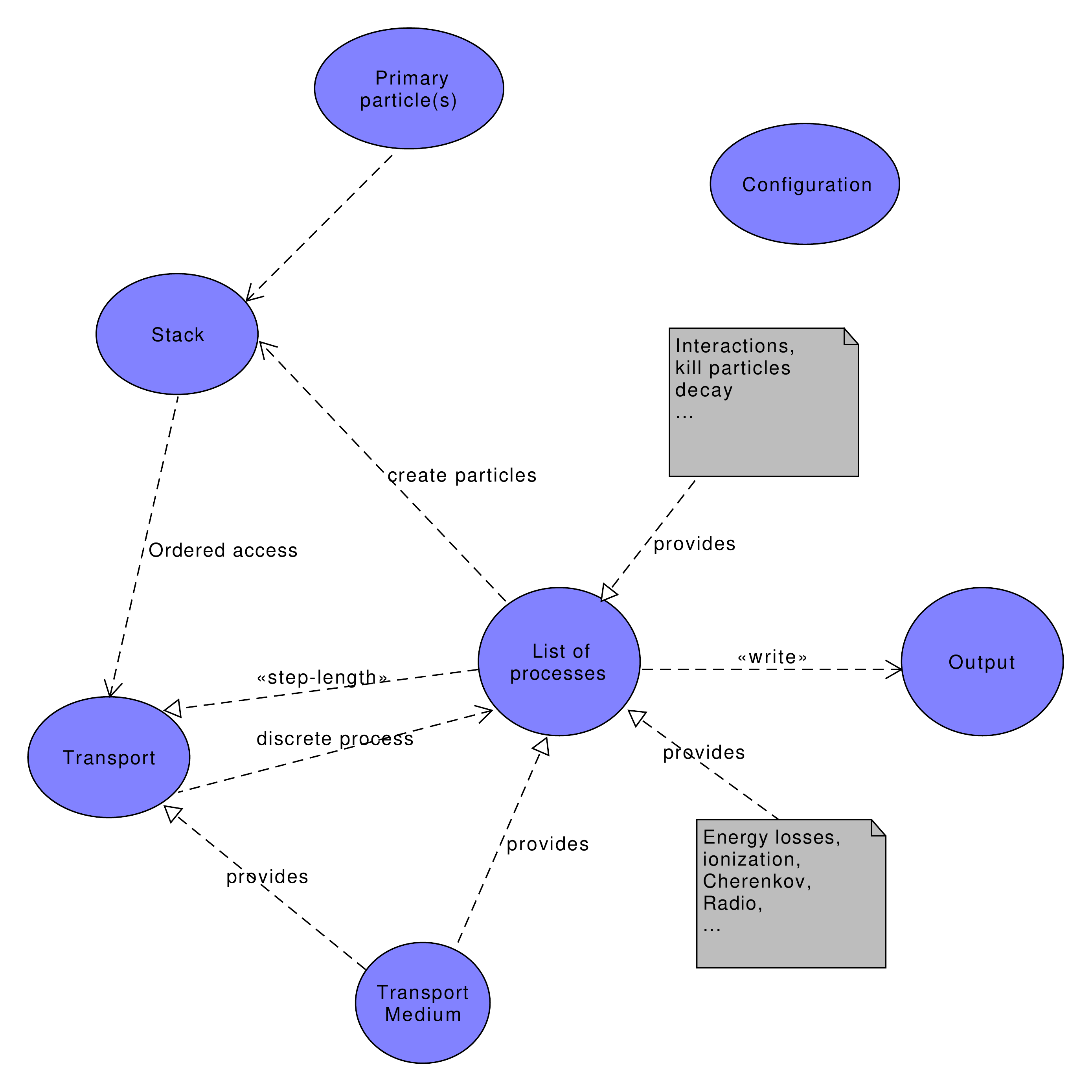}
  \caption{Scheme of particle transport code with processes.}
  \label{fig:simple}
\end{figure}

Some of the goals to achieve with \ngc are extensibility, flexibility,
modularity, scalability, and efficiency.
The main outline of the steps of a typical particle transport code with
processes is illustrated in Fig.~\ref{fig:simple}. The central loop involves a
stack used for temporary storage of particles, a geometric transport code, and a list of
processes that can lead to secondary-particle production or absorption. It is one aim
of the \ngc project to reflect the structural simplicity of Fig.~\ref{fig:simple} to a
very large degree. 

\subsection{Overcoming limitations of current CORSIKA}

The current CORSIKA implementation has a number of limitations, originating
mostly from optimization to specific use-cases as well as the adaption to
more and more novel use-cases which were previously unconsidered in the design
of CORSIKA. These limitations, most of which we intend to remedy in \ngc,
and our anticipated improvements include the following items:
\begin{enumerate}
\item The interaction medium is air with its density being analytically modeled
by five piecewise-exponential layers. \ngc should support arbitrary media 
(air, liquid and frozen water, lunar regolith, rock salt, etc.) and also transitions between 
them. In addition to density, the medium should provide refractive index, 
humidity, temperature, and possibly other information. Medium properties 
might also need to be fed back to the cascade simulation, e.g.\ by 
influencing various energy cutoffs.
\item In CORSIKA, processes taking the cascade simulation as input, e.g.\ radio
or Cherenkov-light calculations, cannot feed back information to the cascade
simulation. In \ngc, any process should be able to give a useful feed-back,
e.g.\ by requiring a change of simulation step-size.
\item In \ngc, an interface should be provided for easy addition of new
interaction models which treat particles or energy ranges not covered by any
other interaction model.
\item CORSIKA does not allow for particle oscillations (e.g.\ neutrinos,
$\textrm{K}^0_\textrm{L}/\textrm{K}^0_\textrm{S}$). A discussion whether or not
oscillations should be incorporated in \ngc should be started.
\item Support for inspecting and storing the history of
  ground-reaching particles (with the EHISTORY
  option~\cite{Heck:2009zza}) is very limited due to rigid memory layout. 
\item No upward-going Cherenkov photons can be handled.
\item It is not envisaged that a started shower simulation is
  \emph{canceled} for any reason, for example when it could be flagged during the simulation process as
  being not relevant for a specific physics study. 
\item Nuclei are supported only up to $Z=60$ and $A=99$.
\item No standardized visualization and validation tools for detailed 
  inspections are provided.
\end{enumerate}

\subsection{Related projects and previous work}

\ngc will heavily depend on expertise gained with the original CORSIKA program.
In addition, experience gained in other projects will be taken into account:
\begin{itemize}
\item MCeq~\cite{Fedynitch:2015zbe} is a recent tool dedicated to the numerical
solution of the CE. It already offers GPU support and very high
computational efficiency. CONEX~\cite{Bergmann:2006yz} is a CE
air-shower simulation program that has been integrated in CORSIKA and provides
enormous increase in computation speed.

\item dynstack~\cite{Baack:2016} is a recent extension of CORSIKA. Its basic
functionality should be adopted for the stack of \ngc.

\item COAST~\cite{coast} has been developed for CORSIKA with the aim of
offering scientists a plugin-like extensibility. The fundamental functionality
of COAST will be available in \ngc.

\item \Offline~\cite{Argiro:2007qg}, the offline analysis framework of the
Pierre Auger Observatory, offers a versatile interface to and implementation
of concepts related to geometry and coordinate systems.

\item Other programs also combine tracking and physics processes, but with
emphasis on different aspects than what is needed in air-shower simulations;
examples are CRPropa~\cite{Armengaud:2006fx} and \textsc{Geant4}~\cite{Agostinelli:2002hh},
although the latter is sometimes used for air showers simulations, too~\cite{Anchordoqui:2000ra,Sanjeewa:2007}.
\end{itemize}

\subsection{Output}

Physics processes of any type can produce various pieces of output information.
These can be particle lists, profiles, histograms, text, etc.  Therefore, when
this output is written to a disk, it has to be stored in a similarly structured
file format, since we require that all the relevant output ends up in one
single file only. However, output can also be written to other places than
disk, for example directly as input for subsequent workflow steps, network
sockets, or other programs -- but this is subject to requirements from the user
community.  The user should be able to decide what kind of output is optimal
for his specific case.

The old binary output format ``DATxxxxxx'' file can still remain as a legacy
option with known limitations. The new standard output, however, will have an
internal directory structure.  Different processes will produce their output in
specific places in this structure. The content of the output file will change
with the choice of the processes and their configurations.

HDF5~\cite{hdf5} is an obvious choice to be considered, while still
having the potential disadvantage of being an external
dependence. ROOT~\cite{Brun:1997pa} could be another possible option.
In any case, we will provide a flexible output interface that physics modules
can rely upon in an implementation-agnostic way.

\subsection{Computational efficiency}

Computational efficiency is not optional for \ngc. The efficient use of
expensive large-scale resources is a crucial requirement, and must be planned
and considered from the early on.
The priority is given to performance over run-time flexibility. The most
fundamental settings of the simulation must be defined at compile time in a
static way: the type of stack, including particle-level data content, physics
models, environmental models, etc. Of course, all models can have additional
parameters that can be defined and modified at run-time.

In general, the use of run-time dynamic design patterns like virtual
classes or dynamic libraries should be minimized (i.e.\ avoidance of virtual
methods in hot code paths). Static design patterns are preferred. 

Data copy operations must be minimized, or performed as late as
possible. The use of ``lazy'' functionality, which is executed only
delayed and when the result is actually needed, should be promoted.

Compiler and CPU optimization should be fully considered for \ngc. Production
versions of the code should claim full benefits from all the available
optimizations.  The execution of particular code on GPUs or other hardware
accelerators (or maybe even more custom hardware) must be transparently
possible.

Parallel and multi-core computations are standard, and are built into
the core of \ngc.

\section{Tools and infrastructure}

The main development infrastructure for \ngc will be provided by our group at KIT. This is
mostly the organization, discussion platform, scientific coordination,
steering, and maintenance of the core functionality. The most useful and
widespread tool for collaborative development available today is the
version control system is git. Git allows having a very
dynamic and large base of contributors, and at the same time a well controlled
access to the main code-base via \textit{pull requests} (PRs). The code review,
discussion, testing and validation of PRs will be an important task of the
project
steering. Code will be peer-reviewed, with an emphasis on clearness and
readability, and inline documentation (doxygen).  Furthermore, automatic
unit-testing and validation will be performed.
Unit tests must yield a very high coverage of the \ngc code. Unit tests are
executed automatically by a jenkins (or equivalent) service to perform
low-level code and PR validation.
Additional automatic validation and high-level tests must accompany the regular
testing, cover all the important functionality and, in particular, all physics.

Automatic testing will provide a well defined list of supported environments,
combined with a control over a specified set of different selections of
simulation options.

We use the gitlab server \texttt{https://gitlab.ikp.kit.edu} for the hosting.
This gitlab server provides also an
issue tracking functionality that is linked to defined
milestones. A wiki page service is also provided. Connect to this
server to see the status of the \ngc project, download releases, or
even get directly involved in discussions or the development.

\section{Main challenges}

While there are many challenges to overcome, a list of topics that require
particularly dedicated attention is given in the following. These topics are more-or-less
directly linked to the underlying/internal physics of the cascade process and
require very intelligent and likely highly-complex solution.
\begin{enumerate}
\item Efficient integration of electron-gamma cascades (previously EGS4).

\item Random number generation in an inherently multi-core and parallel
environment while ensuring the full reproducibility of simulations.

\item Investigating the limits of equivalence between CE-solving and
detailed Monte Carlo transport methods (d$E$/d$X$, Cherenkov, lateral structure, radio
production, etc.).

\item GPU optimization.

\item Scalability in supercomputing environments.
  
\end{enumerate}

\section{Details}

\begin{figure}
\centering
\includegraphics[width=\textwidth]{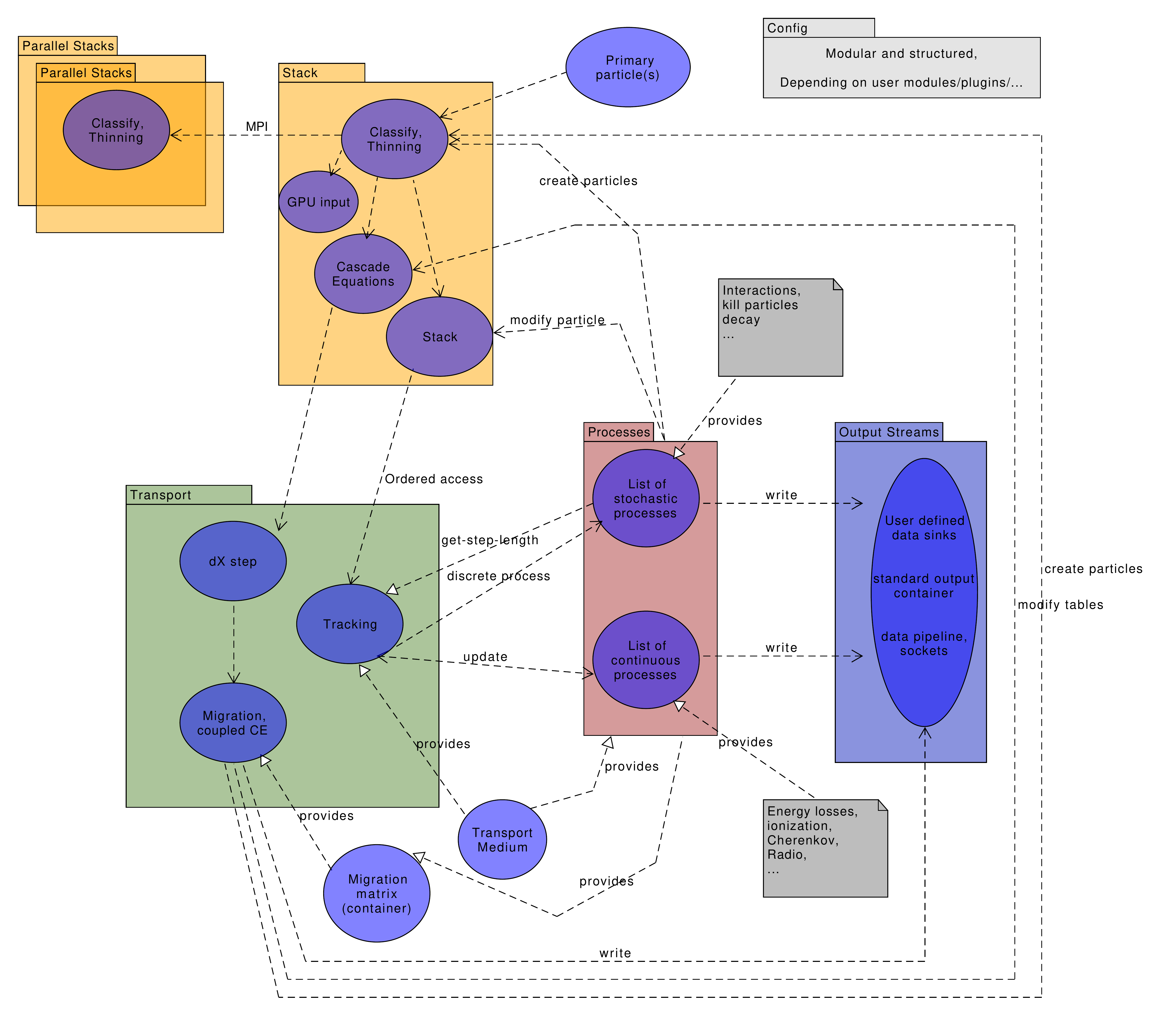}
\caption{Main building blocks and workflow steps of \ngc which already highlight
the fundamental functionality and flexibility.}
\label{fig:scheme}
\end{figure}

Taking the aforementioned considerations and requirements into account, a more
detailed scheme of the simulation workflow becomes necessary, as outlined
in Fig.~\ref{fig:scheme}.  Some of the aspects of this diagram still need to be
optimized or determined precisely.  Nevertheless, with the basic design as
given here, the modular functionality and building blocks can be developed in
parallel. Rudimentary definitions of interfaces needed for these purposes are
given below.

Note that the code fragments given as examples here are, first of all, not in
any specific language and do not follow any specific syntax. This is a pure
pseudo-code used to illustrate the basic functionality and employed patterns,
and only vaguely resembles C++.

\subsection{Conventions and coding}

A programming language offering high level of design flexibility and at the same
time excellent compiler and optimization support is required. It is an
advantage to chose a language that also has non-science relevance and thus
assures long-term support, development, and expertise.
For this purpose, we decided to use C++. At the beginning \ngc will be based on
the C++17 standard, a choice that will most probably evolve in the future.

General guidelines for contributing of the code will be well defined and must
be strictly enforced~\cite{isocpp}.  These guidelines will be distributed via
the documentation section on the gitlab server mentioned above and/or the wiki pages.
The guidelines can be discussed, agreed upon, and also improved in discussions
between the developers and the project steering.  One of the most important
things in such a project is communication -- and the code will be the prime means
of communication between the team members~\cite{Zakas:2012}, since, let us not
forget, most of the time people will spend on this project will be dedicated
to reading other people's code~\cite{martin2009clean}.  A more exhaustive list
of core guidelines for C++ can be found in Ref.~\cite{CppCore}.
Those items are also relevant in this respect:

\begin{itemize}
\item Code must be accompanied with inline comments. Note that a well chosen
naming of identifiers and functions can greatly reduce the burden of
documenting the code. A well written code is self-explanatory to a large extent. In addition, for
systematic documentation doxygen commands must be used where possible.
  
\item One aspect of choices of the style should be to minimize the probability
of programming errors. For example pointers should be used only where
absolutely necessary, and that should never be exposed to the user. 

\item We will favour static over dynamic polymorphism. On a low level of the
code, this will lead to the abundant use of templates. However, high-level
users and physicists should not be exposed to templates, unless absolutely
required. 

\item Test-driven development is encouraged. Therefore, from early on, a useful
setup of unit tests should be supported by the build system.  The unit testing
will be an essential part of \ngc. A high coverage of code by tests will be a
prime criterion for acceptance. 
\end{itemize}

\subsection{Dependencies}

The use of external code and libraries must be minimized to the absolute
minimum in order to stay conflict-free and operational over a very extended
period of time. Individual exceptions might be possible, but must be well
motivated and discussed before getting included into the mainline code.
For each functionality we should evaluate whether a basic re-implementation is
more feasible than inclusion of an external dependency. In any case, whenever possible, appropriate wrappers in \ngc should
hide the implementation details of external packages in order to keep replacement or re-implementation option open without
a need for breaking the interface.
Likely packages and options for external libraries are (excluding
packages that will be distributed together with \ngc):
\begin{itemize}
\item C++17 compiler.
\item CMake build system.
\item git [for development].
\item doxygen [for development].
\item presumably \texttt{boost} for \texttt{yaml} and \texttt{xml},
	histograms, file system access, command-line options, light-weight configuration parsers (property tree), random numbers, etc.
\item HDF5 and/or ROOT for data storage [at least one of both required].
\item PyBind11~\cite{pybind11} for bindings to Python.
\item HepMC~\cite{Dobbs:2001ck} as generic interface, also for exotics [optional].
\item
In order to generate random numbers, we will
use standardized interfaces and established methods.
For testing
purposes, the possibility to exchange the random-number engine should be relatively easy.
No homegrown generators and only well established, checked, and vetted
methods for generating random numbers should be used, likely provided
by \texttt{boost} as well.
\end{itemize}

Lightweight packages like small header-only libraries can be distributed together
with \ngc. Likely candidates are:
\begin{itemize}
\item Eigen3~\cite{eigen} for linear algebra.
\item catch2~\cite{catch2} for unit tests.
\item PhysUnits~\cite{PhysUnits} for units (see below).
\end{itemize}

\subsection{Configuration}

The framework has to support extensive run-time (from configuration
files or on command line) as well as compile-time configuration. The
latter involves conditional compilation, static polymorphism,
switching between policies in policy-driven classes. 

The run-time configuration will support structured yaml or xml as input, either
in a single file, or multiple files located in a directory. Modules of \ngc can
retrieve the required configuration via a global object in a
structured way. Command-line options are parsed and provided via the
same mechanism. By default, the complete configuration will be saved
into the output file, and will thus, if needed, allow identical reproduction of a
simulation at a later time. Physics modules can access configuration
via a \emph{section name} and a \emph{parameter name}, for example
\begin{verbatim}
  primaryEnergy = Config.Get("PrimaryParticle/Energy");
\end{verbatim}
\noindent
where \texttt{PrimaryParticle} is the name of the configuration
section, and \texttt{Energy} the parameter. The data can be obtained from
files, or provided via the command line,
for example via \texttt{--set PrimaryParticle/Energy=1e18\textunderscore eV}.

For more intricate situations where a simple configuration file might
not be sufficient, or when a dynamic change of parameters during runtime is needed,
the simulation process can be more conveniently steered by means of a
script. The library \textit{PyBind11} allows us to provide bindings
to Python with minimal efforts. 

\subsection{Units}

\ngc will utilize the header-only library \textit{PhysUnits} for
handling quantities having physical dimensions (i.e. "units"). First, it allows
us to conveniently attach units to the numerical literals in the code (e.g.\ 
\verb|auto criticalEnergy = 87_MeV;|), thereby avoiding other, hard to enforce
explicit conventions and improving readability, especially in a collaborative
environment.

Second, as the dimensions of quantities are encoded in their respective types, a dimensional analysis is
imposed upon computations involving dimensionful quantities during the compilation.
This way, an otherwise silent error of mismatched units is converted to a compile-time error,
as in the following example:
\begin{verbatim}
  Length_t distance = 47.2_cm;
  Time_t time = 35.9_ns;
  Speed_t speed = distance + time; // compiler error!
  Frequency_t freq = 1 / distance; // compiler error!
\end{verbatim}
\noindent
During compilation the conversion of quantities to common base units (which
the developer does not need to know and are internally chosen to minimize numerical errors) is performed.

Because of this functionality, this approach is more restrictive than more simplistic implementations
like e.g.\ provided in \textsc{Geant4}/CLHEP~\cite{Agostinelli:2002hh,Lonnblad:1994np}, where units are
provided only as a set of self-consistent multiplication constants. We believe, nevertheless,
that the use of "strongly-typed units" will make development less error-prone.

At the same time, no runtime overhead is introduced when compiler optimizations are
enabled since, after all, such a dimensionful quantity in memory is just the usual
floating-point number.

\subsection{Geometry, coordinate systems and transformations}

A key ingredient to the usability of \ngc is the ability to conveniently
work with geometrical objects such as points, vectors, trajectories, etc.,
possibly defined in different coordinate systems.
We will provide a geometry framework (with unit support fully integrated),
to a large extent inspired by \Offline, in which geometrical objects
are defined always with a reference to a specific coordinate system.
In our case the relevant coordinate systems mainly comprise the environmental
reference frame and the shower frame, but additonal systems can be defined
as needed.  When dealing with multiple objects at the same time, e.g.
\verb|sphere.IsInside(point)|, is it automatically taken care of transforming
the affected objects into a common reference frame. Therefore, when one
can formulate his computations in a way that does not involve any specific
coordinate system, the handling of potentially necessary transformations stays
completely transparent.

As possible transformations that define coordinate systems with respect
to each other we restrict ourselves to elements of the special Euclidean
group \textsl{SE}$(3)$ (see e.g.\ ref.~\cite{Ivancevic:2011vc}), i.e.\ rotations
and translations. 
Although one might favor Poincaré transformations as they include Lorentz
boosts, which are certainly required for interfacing external interaction
models, this would require to add a time-like coordinate to all geometric objects.
This adds significant complexity to the code in our setup that is otherwise completely 
static. 
For example, the concept of a point fixed in space in the lab frame would
require to be upgraded to a world line. We currently do not envisage to support
modeling of relativistic moving objects in our environment -- except for the particles,
of course -- as this would significantly complicate and slow down our particle
tracking algorithms.
Due to the special properties of rotations and translations it is not computationally
expensive to perform inverse transformations because expensive matrix inversions
can be avoided.

Regarding the aforementioned Lorentz boosts, special attention must be paid
to ensure numerically accurate results in all relevant regimes, comprising the range
from non-relativistic ($\beta \ll 1, \gamma \simeq 1$) to ultra-relativistic
($\beta \simeq 1, \gamma \gg 1$) boosts.

\subsection{Particle representation}

The typical minimal set of information to describe a particle is:
type, mass, energy-momentum, and space-time position.  In certain use cases
this can be extended for example with (multiple) weights, history
information (unique ID, generation, grandparents, interaction ID), or
further information.

Interaction models typically do not care about the space-time part since once the model is
invoked according to the total cross-section, the impact parameter is determined
internally by the model in a small Monte Carlo procedure (and not e.g.\ from the microscopic positions of air
nuclei in the atmosphere).  Nevertheless, the propagation and the continuous
losses will eventually need the space-time parts of the particle information.

Particle properties like mass and lifetime are extracted from
the \texttt{ParticleData.xml} file provided by PYTHIA\,8~\cite{Sjostrand:2014zea},
together with their PDG code~\cite{Tanabashi:2018oca}. To allow for
efficient lookup of these properties, the \ngc-internal particle code is chosen to be
different than the PDG code. Since the PDG codes only very sparsely cover a large
integer range, they are not very useful as indices in a lookup table. \ngc therefore
uses a contiguous range of integers which is automatically generated from
the union of all particles known by the user-enabled interaction models. Rather
than using these integers directly in the \ngc code, however, \texttt{enum}
declarations will be provided for convenience and improved code readability.
In contrast to their corresponding numerical values, the \texttt{enum}
identifiers (e.g.\ \verb|Code::DStarMinus|) are guaranteed to be stable after 
recompilation with different interaction modules, as well as in future
\ngc releases.

For this purpose, the needed code is generated by a provided script
before the actual compilation of \ngc. This script will depend on the aforementioned file from
PYTHIA. The output is C++ code that will allow to write
expressions like these:
\begin{verbatim}
  // compile-time evaluated expressions:
  auto constexpr mElectron = ParticleData::GetMass(Code::Electron);
  auto constexpr tauPi = ParticleData::GetLifetime(Code::PiPlus);
  ...
  // run-time evalueted expressions:
  auto particleType = stack.GetNextParticle().GetType();
  auto charge = ParticleData::GetCharge(particleType);
\end{verbatim}
%
\noindent
The internal numeric particle-ID is just an index, the representation of
particles in \ngc code and \texttt{enums} is obtained from the
particle names in the \texttt{xml} file.
When specific interaction models internally use different schemes for
particle identification, extra code is provided in the interface part
to those models, where the conversion between the external and internal codes is
performed.

For binary output purposes, however, \ngc-internal codes are converted
to the well known, standardized PDG codes to ensure seamless interoperability
with other software packages used within the HEP community.
In any text output, e.g.\ log files, the output is by default converted to a
human-readable identifiers. For example, \verb|cout << someParticleCode << endl;|
might, depending on the value of \verb|someParticleCode|, print out ``\texttt{e-}'' or ``\texttt{D+}''
unless a numerical output (in \ngc or PDG scheme) is explicitly requested.

\subsection{Framework}

The \ngc consists of an inner core and associated modules that can
also be entirely external. Thus, there can be -- and generally is -- a
distinction between code in the ``core'' of \ngc and ``outside'' of this, defining a ``frontier''
where conventions, units, and all kind of reference frames have to be
adapted and converted in a consistent way. Most obviously is this the case for all
existing hadronic event-generators and input/output facilities. Nevertheless, this can occur also in other components,
and the frontier can thus occur at different places.  The code needed for
the conversions in the frontier must be provided together with the
\ngc framework. 
Special care must be taken in
cases where different models, for example, use different constants for the mass
of particles, which can lead to numerically unreasonable results like negative
kinetic energies or invalid transformations. The details of such effects must be investigated and a
comprehensive solution has to be found at a later time.

\subsubsection{Particle processing and stacks}

A core concept of \ngc is that particles are stored on a dedicated stack. This
is needed since in cascade processes an enormous number of particles can
be accumulated, requiring careful handling of such data. The stack can
automatically swap to disk when memory is exhausted. The access and
handling of particles on the stack has an important impact on the performance
of the simulation process. In typical applications it is optimal in terms of
memory footprint to process the
lowest-energy particles first, but there can be situations where completely
different strategy becomes relevant. The stack should be flexible enough to allow
various user-specific interventions, while the simulation is writing to and reading from it.

In \ngc there is no need to have a dedicated persistent object describing
an individual particle. Particles are always represented by a reference/proxy
to the data on the stack. On a fundamental level, such stacks can be a FORTRAN common
block, dynamically-allocated C++ data, a swap file, or any other source/storage of
particle data.

\subsubsection{Main loop, simulation steps, processes}

A central part of \ngc is the loop over all particles on the stack. These particles are
transported and processed in interactions with the medium, and as part of this also CE
tables can be filled.  All these processes can produce new particles or
modify existing particles on the stack. Furthermore, the processes can produce
various output data of the simulation process. CE migration-matrices are
either computed at program start or read from pre-calculated files. When the
stack is empty (or any other trigger), the CE
are solved numerically, which can, once more, also fill the particle stack.
Thus, a double-loop is required here in order to process the full particle
cascade:
\begin{verbatim}
  while (!stack.Empty()) {
    while (!stack.Empty()) {
      auto particle = stack.GetNextParticle();
      Step(particle);
    }
    cascadeEquations.Solve();
  }
\end{verbatim}
\noindent
The transport procedure needs to handle geometric propagation of neutral and charged
particles, thus, magnetic and electric deflections are important.  The
transport step-length is used to distinguish two type of processes:
\begin{itemize}
\item \emph{Continuous} processes occur on a scale much below the transport
step-length, e.g.\ ionization, and thus an effective treatment can be used.

\item \emph{Discrete} processes typically lead to the disappearance of a
particle and to production of new particles (typically in, but not limited to, collisions or decays).
\end{itemize}
The optimal size of the simulation step is determined from the list of all processes
considered. The discrete process with the highest cross-section limits the
maximum step size. However, also a continuous process can limit the step size,
for example by the requirement that ionization energy-loss, the multiple-scattering
angle, or the number of emitted Cherenkov photons cannot exceed
specific limits.  Furthermore, even particle transport is just a specific type
of process which propagates particles. Since the propagation can lead a particle from one
medium (e.g.\ the atmosphere) into another (e.g.\ ice), the particle transport
can also have a limiting effect on the maximum step length allowed. An individual step cannot cross
from one medium to another, but for correct treatment must terminate at the boundary between the two
media.  Furthermore, the particle transport in magnetic
fields leads to deflections, where step size has to be adjusted according to the curvature of the deflection.

Thus, the geometric particle-transport must be the first process to be executed.
The information about the particle trajectory is important input for the
calculation of subsequent continuous processes. Finally, the type and
probability of one single discrete process is last to be determined for each
simulated transport step.  The simulated discrete process is randomly selected,
typically according to its cross section or lifetime. The structure of
the code to execute in one simulation step is thus:
\begin{verbatim}
  Step(Particle& particle)
  {
    auto stepLength = MaximalStepLength(tracking, continuousProcesses,
                                        stochasticProcesses);
    auto trajectory = tracking.Propagate(particle, stepLength);
    for (auto& cp : continuousProcesses) {
      cp.Propagate(particle, trajectory, stepLength);
    }
    // randomly select ONE or NONE stachastic process
    if (discreteProcess dp = SelectStochasticProcess(stepLength)) {
      dp.Interact(particle);
    }
  }
\end{verbatim}
\noindent
The numerical solution of the CE is performed as being functionally fully equivalent
to a normal propagation. While some of the processes can easily be
formulated using migration matrices, our aim is, though, to scientifically
evaluate and exploit the concept as extensively as possible, covering the
production of Cherenkov photons, radio emission, etc. The data for the CE is
stored in a \emph{table} (which in general will cover multiple dimensions)
representing histograms, for example of the number of particles of specific type
versus energy. The \emph{migration} of particles to different bins in energy
\emph{and} to different particle types is described by pre-computed migration-matrices.
The matrices implicitly already encode the information on the geometric
length of simulation steps. In some aspects the CE approach corresponds to the
approximation where the discrete processes are handled like continuous
processes. This is reflected in the structure of the corresponding code:
\begin{verbatim}
  CascadeEquations::Solve()
  {
    while (!table.Empty()) {
      for (auto cp : continuousProcesses) {
        cp.CascadeEquationPropagate(table)
      }
      for (auto dp : discreteProcesses) {
        dp.CascadeEquationPropagate(table)
      }
    }
  }
\end{verbatim}
The limits of the application of CE to specific processes are not known
precisely at this moment and certainly there are various challenges facing us ahead.
Particularly difficult processes are those which depend significantly on geometry, like
Cherenkov or radio emission. It is up to the detailed studies to evaluate their performance and
adapt the methods to potential (limited) use cases.  This will be subject of research as
part of the project.

\subsection{Radio}

Radio-emission calculations, which in the original CORSIKA are provided by the CoREAS extension \cite{HuegeARENA2012a},
rely on the position and timing information 
of charged particles to calculate the electromagnetic radiation 
associated with a particle shower. With its increased flexibility, \ngc 
will enable radio-emission calculations for a much larger range of 
problems. In particular, simulation of the radiation associated with 
showers penetrating from air into a dense medium or vice-versa will 
become possible due to the more generic configuration of the interaction media.  
Feedback of the radio calculation to the cascade simulation (e.g., 
modifying simulation step-sizes or possibly thinning levels) might 
increase performance and/or simulation accuracy. GPU parallelization has the
potential to greatly reduce computation times, which are currently the main bottleneck
for simulations of signals in dense antenna arrays. Simulations in media 
with a sizable refractive-index gradient will require certain ray-tracing 
functionalities, possibly even finite-difference time-domain calculations. The modular approach of 
\ngc will allow the implementation of different radio-emission 
calculation-formalisms and enable systematic studies of their 
differences.

\subsection{Environment}\label{sec:environment}

Traditionally the medium of transport for CORSIKA was the Earth's
atmosphere. It is one of the purposes of \ngc to allow for much more flexible
combination of environments. This includes water, ice, mountains, the
moon, planets, stars, space, etc.  In this case also the interface
between different media becomes a matter of significance for the
simulation.  
Showers can start in one medium and subsequently traverse into different media.
The environment will be a dedicated object to configure
for every physics application. The structure of the environment will
be defined before compilation, the properties of the environment can
be configured via configuration files in any way needed for the
application. This can be either static, or time-dependent.

The global reference frame is specified by the user and depends on the
chosen environmental model. For a standard curved Earth this is the
center-of-the-earth frame. With double floating-point precision this yields a precision better
than a nanometer over more than $10\,000$\,km distance.

Particles are tracked in the global reference frame. The secondary particles produced by
discrete processes occurring at specific locations in the cascade are
transformed and boosted back into the global coordinate frame.

For specific purposes, like tabulations and some approximations, the \emph{shower
coordinate-system}, in which the $z$-axis points along the primary-particle momentum,
can also be relevant.

The initial randomization of primary-particle locations and directions is
performed by dedicated modules, which can be changed and configured by the users
to get on the detector level the desired distributions.
The environment object provides all of the required access to the environmental
parameters, e.g.\ roughly in the following form:
\begin{verbatim}
  Environment::GetVolumeId(point)
  Environment::GetVolumeBoundary(trajectory)
  Environment::GetTargetParticle(point)
  Environment::GetDensity(point)
  Environment::GetIntegratedDensity(trajectory)
  Environment::GetRefractiveIndex(point)
  Environment::GetTemperature(point)
  Environment::GetHumidity(point)
  Environment::GetMagneticField(point)
  Environment::GetElectricField(point)
\end{verbatim}
This interface is sufficient since, for example a concept like altitude, defined as
distance from a point to a surface on a direct line to the origin (center of
the Earth), is needed only internally within the environment object.

The environment object will use a C++ policy concept to provide access to the
underlying models. This requires re-compilation after changes in the model setup.
However, individual models can still be configured at runtime.

\subsubsection{Geometric objects}

We will keep the geometry description as simple as possible and to the level
needed to achieve the physics goals. At the moment, this goals include
being able to define different (typically very large) environment regions with
distinct properties. Initially, it is sufficient to provide only
the most simple forms and shapes, e.g.\ sphere, cuboid, cylinder, and maybe
trapezoid as well as pyramid. The geometry package must be structured in a
generic way, so that it can be extended, if needed, to include more
complex and fine-grained objects at a later time.
We are not planning to support general-purpose geometry as, for example
in \textsc{Geant4}~\cite{Agostinelli:2002hh}. When in a specific volume
of the simulation a very complex geometry is required, it is probably
the best choice to allow seamless integration of \ngc with
\textsc{Geant4}, where particles can be passed-on from one package to the
other.

\section{Summary}

The steps towards creation of \ngc outlined here are optimized to best support 
scientific research in fields where the simulation involves particle transport and particle
cascades with stochastic and continuous processes. The targeted goals of
the resulting framework will be far beyond the capabilities of the original CORSIKA program. It is
up to the scientific community to decide in which concrete applications \ngc
will be used in the future. It is our aim to offer long-term support for the
\ngc program over a period of more than 20\,years.

The modularity of the proposed code and the magnitude of the project offers the
opportunity for the scientific community to participate in a collaborative manner.
Specific functionality and modules can be provided and maintained by different
groups. The core of the project, the integration, and the steering are provided by KIT. This
can be also a suitable model for a scenario where different communities have
different requirements, but the overall collaborative approach is the one we want to promote and foster.
This will require dedicated and strict commitment to the project from all the participating parties in order to
assure the stability and functionality with no compromises needed.

A better access to the air-shower physics-simulation process will be one of the
keys to address the main open questions of cosmic-ray physics, the universe at
the highest energies, and related scientific problems.

\section*{Acknowledgements}

We thank the participants of the Next-Generation CORSIKA Workshop for their
valuable comments and suggestions regarding this white paper and the future
of CORSIKA.
T.H.\ acknowledges very fruitful discussions within the radio-detection community.
M.R.\ acknowledges support by the DFG-funded Doctoral School ``Karlsruhe
School of Elementary and Astroparticle Physics: Science and Technology''.

\printbibliography 

\end{document}